\documentclass[preprint,10pt,5p]{elsarticle}

\usepackage[cmex10]{amsmath}
\usepackage{url}
\usepackage{verbatim}
\usepackage{cancel}

\PassOptionsToPackage{hyphens}{url}
\usepackage{hyperref}

\DeclareGraphicsExtensions{.pdf}
\usepackage[caption=false,font=footnotesize]{subfig}

\usepackage[table]{xcolor}
\usepackage{multirow}
\colorlet{yes}{green!20}
\colorlet{no}{red!20}
\colorlet{maybe}{orange!20}

\newcommand{\no}{\cellcolor{no}No}
\newcommand{\yes}{\cellcolor{yes}Yes}
\newcommand{\maybe}[1]{\cellcolor{maybe}#1}
\newcommand{\bad}[1]{\cellcolor{no}#1}
\newcommand{\good}[1]{\cellcolor{yes}#1}

\newcounter{IoTPlatform}
\newcommand{\iotplatform}[2]{
\vspace{1em}\par\noindent\refstepcounter{IoTPlatform}Platform
\arabic{IoTPlatform}: \textbf{#1} ({\scriptsize\url{#2}})\par}
\newcommand{\citeiot}[1]{[Platform~\ref{#1}]}

\usepackage{tkz-kiviat} 
\usetikzlibrary{arrows}

\definecolor{c1}{RGB}{0,0,100}
\definecolor{c2}{RGB}{43,0,255}
\definecolor{c3}{RGB}{199,43,230}
\definecolor{c4}{RGB}{255,141,115}

\journal{Computer Communications, Special issue on the Internet of
  Things: Research challenges and Solutions}

\begin{document}

\begin{frontmatter}
  \title{A gap analysis of Internet-of-Things platforms}
  \author[uh]{Julien~Mineraud\corref{cor1}}
  \ead{julien.mineraud@cs.helsinki.fi}
  \author[jyy]{Oleksiy~Mazhelis}
  \ead{mazhelis@jyu.fi}
  \author[oy]{Xiang~Su}
  \ead{xiang.su@ee.oulu.fi}
  \author[uh]{Sasu~Tarkoma}
  \ead{sasu.tarkoma@cs.helsinki.fi}

  \cortext[cor1]{Corresponding author}
  \address[uh]{Department of Computer Science, University of Helsinki,
    Finland}
  \address[jyy]{Department of Computer Science and Information
    Systems, University of Jyv\"{a}skyl\"{a}, Finland}
  \address[oy]{Center for Ubiquitous Computing, University of Oulu, Finland}
  
  \begin{abstract} 
    We are experiencing an abundance of Internet-of-Things (IoT) middleware 
    solutions that provide connectivity for sensors and actuators to the Internet.
    To gain a widespread adoption, these middleware solutions, referred to as 
    platforms, have to meet the expectations of different players in the IoT
    ecosystem, including device providers, application developers,
    and end-users, among others.

    In this article, we evaluate a representative sample of these platforms, 
    both proprietary and open-source, on the basis of their ability to meet 
    the expectations of different IoT users. 
    The evaluation is thus more focused on how ready and usable these platforms
    are for IoT ecosystem players, rather than on the peculiarities of the 
    underlying technological layers.
    The evaluation is carried out as a gap analysis of the current IoT landscape 
    with respect to 
    (i) the support for heterogeneous sensing and actuating technologies,
    (ii) the data ownership and its implications for security and privacy,
    (iii) data processing and data sharing capabilities,
    (iv) the support offered to application developers, 
    (v) the completeness of an IoT ecosystem, and
    (vi) the availability of dedicated IoT marketplaces.
    The gap analysis aims to highlight the deficiencies of today's solutions 
    to improve their integration to tomorrow's ecosystems. 
    In order to strengthen the finding of our analysis, we conducted a
    survey among the partners of the Finnish IoT program,
    counting over 350 experts, to evaluate the most critical issues for the
    development of future IoT platforms.
    Based on the results of our analysis and our survey, we conclude this
    article with a list of recommendations for extending these IoT
    platforms in order to fill in the gaps.
  \end{abstract}


\end{frontmatter}

\section{Introduction}\label{sec:introduction}

The Internet of Things (IoT) paradigm foresees the
development of our current environment towards new enriched spaces,
such as smart cities, smart homes, smart grid, digital health, and automated environmental pollution
control~\cite{Borgia2014,Atzori2010}.

In recent years, an abundance of solutions has emerged to interconnect
smart objects for systems with different scales and objectives. 
For instance, a lightweight platform can be deployed in one's home to 
orchestrate several connected objects, such as the fridge, the lights, and the heating system. 
On a broader scale, a smart city may benefit its development and
management from new IoT solutions that can handle thousands of
sensors, ease their maintenance, recalibration and, more importantly,
analyze the data that they produce~\cite{Lea2014a,Tsai2014}.

\begin{figure}[tb]
 \centering
  \subfloat[Cloud-based platform]{\label{fig:cloud_platform}
    \includegraphics[width=.48\linewidth]
    {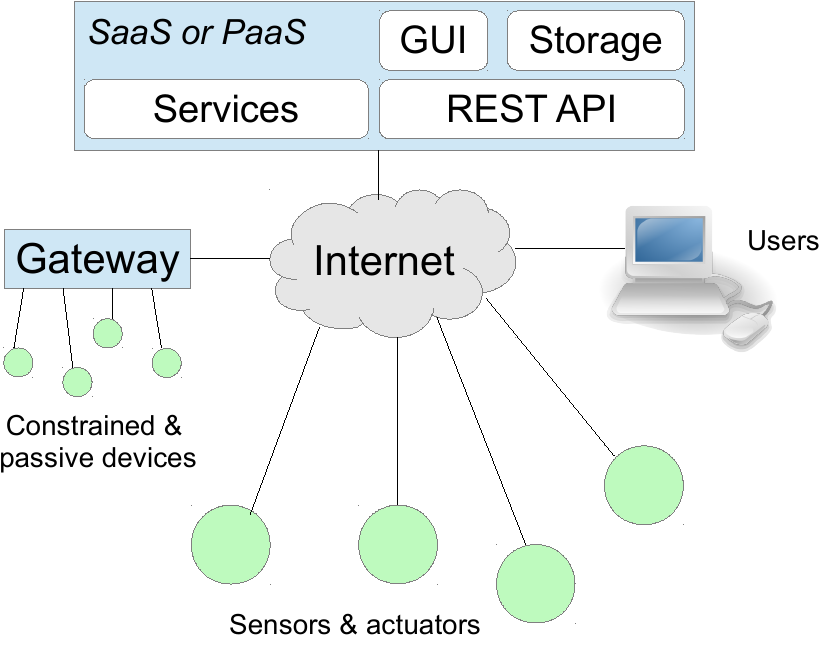}}\hfill
   \subfloat[Local platform]{\label{fig:local_platform}
    \includegraphics[width=.48\linewidth]
    {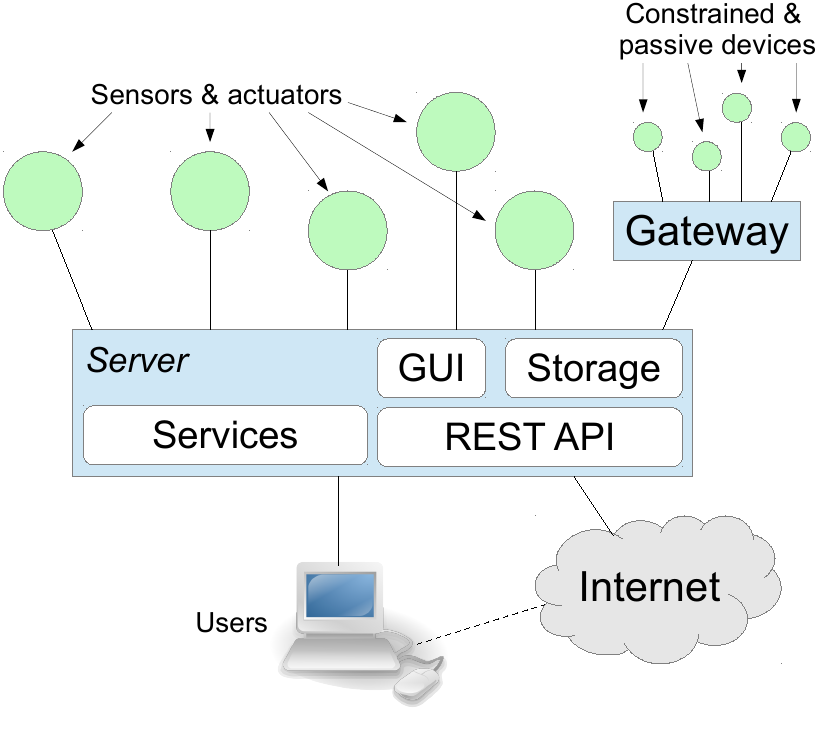}}\\
 \caption{
 	\textbf{End-to-end interactions between users, smart devices and the platform.}
 	\textsl{IoT platforms currently enables interaction between devices and users. 
    However, interactions between IoT platforms are limited and costly.}
 }
 \label{fig:siloed_architectures}
\end{figure}

In this article, we study today's IoT landscape with regard to the
distribution of applications and services, as well as the platforms
that connect the devices to the Internet. For the purposes of this
paper, an IoT platform is defined as the middleware and the
infrastructure that enables the end-users to interact with smart
objects, as depicted in \figurename{~\ref{fig:siloed_architectures}}.
We frame our study as a gap analysis of these platforms with regard to
their capacities in meeting the challenges emerging from the current
development of the IoT technologies. 
In order to evaluate the limitations of the current IoT platform
landscape and identify the gaps that need to be filled, we consider
the viewpoints of different players of the IoT platform ecosystem,
including device vendors, application developers, providers of
platforms and related services, and the end-users.
In order to strengthen the findings of the gap analysis, we conducted
a survey among the experts of the national Finnish IoT
program~\cite{Tarkoma2013} to highlight the most critical gaps for the
development of future IoT platforms.
As a result of this evaluation, we propose a set of recommendations
aimed at filling in the identified gaps.

The remainder of this article is organized as follows:
Section~\ref{sec:review} presents the review of a representative list
of IoT platforms. 
This is followed by a thorough gap analysis of the solutions in
Section~\ref{sec:gap}. 
In Section~\ref{sec:survey}, we present the results of the survey and
in Section~\ref{sec:recommendations}, we enumerate our recommendations
for filling in the gaps outlined in the previous sections. 
Finally, we will conclude this article in
Section~\ref{sec:conclusion}.

\section{Review of today's IoT platforms}\label{sec:review}

In this section, we survey available IoT platforms, both
proprietary and open-source, that connect
\emph{smart objects} or \emph{things} to the Internet.
The list of the 39 platforms being surveyed, ordered alphabetically
and numbered (e.g., ~\citeiot{iot:linksmart}, where
\ref{iot:linksmart} also refers to the ``ref'' column in
Table~\ref{table:IoTplatforms}), 
along with further details about these platforms, can be found in
\ref{appendix}.
The list of the surveyed platforms shall by no means be seen as
exhaustive though we believe that a representative
sample of the available platforms has been included in the survey.

\begin{table*}[!ht]
  \renewcommand{\arraystretch}{1.22}
  \centering
  \caption{Available IoT platforms}
  \label{table:IoTplatforms}
  \scriptsize
  \begin{tabular}{|r|p{7em}|p{7em}|l|l|p{8.5em}|l|l|l|}\hline
  Ref & Platforms & a) Support of & b) Type & c) Architecture & d)
  Open source & e) REST & f) Data access & g) Service\\
  & & heterogeneous devices & & & & & control &
  discovery\\\hline\hline
 
  \ref{iot:airvantage} & AirVantage\textsuperscript{\texttrademark} &
  \bad{Needs gateway} & M2M PaaS & Cloud-based & \maybe{Libraries only
  (Apache v2, MIT and Eclipse v1.0} & \yes & OAuth2 & \no\\\hline
  
 \ref{iot:arkessa} & Arkessa & \yes & M2M PaaS & Cloud-based & \no &
 n.a.
 & \maybe{Facebook like} & \no\\
 & & \good &  & & \bad &  & \maybe{privacy settings} & \bad \\\hline
 
 \ref{iot:armmbed} & ARM mbed & \maybe{Embedded} & M2M PaaS &
 Centralized/ & \no & \good{CoAP} & \good{User's choice} & \no\\
 & & \maybe{devices} & & Cloud-based & \bad{} & \good &
 \good & \bad\\\hline
  
 \ref{iot:carriots} & Carriots\textsuperscript{\textregistered} & \yes
 & PaaS & Cloud-based & \no & \yes & Secured access & \no\\\hline
  
 \ref{iot:devicecloud} & DeviceCloud & \yes & PaaS & Cloud-based & \no
 & \yes & n.a. & \no\\\hline
 
 \ref{iot:everyaware} & EveryAware & \yes & Server & Centralized & \no
 & \yes & \good{4 levels} & \no\\\hline
 
 \ref{iot:everyware} & Everyware & \bad{Needs gateway} & PaaS &
 Cloud-based & \no & \yes & n.a. & \no\\\hline
 
 \ref{iot:evrythng} & EvryThng & \yes & M2M SaaS & Centralized & \no &
 \yes & \maybe{Fine-grained} & \no\\\hline
  
 \ref{iot:exosite} & Exosite & \yes & PaaS & Cloud-based &
 \maybe{Libraries only (BSD license)} & \yes & n.a. & \no\\\hline
 
 \ref{iot:fosstrack} & Fosstrack & \bad{RFID} & Server & Centralized &
 \no & \no & \good{Locally stored} & \no\\\hline
  
 \ref{iot:grovestreams} & GroveStreams & \no & PaaS & Cloud-based &
 \no & \yes & Role-based & \no\\\hline
 
 \ref{iot:hat} & H.A.T. & \maybe{Home devices} & PaaS & Decentralized
 & \yes & \yes & \good{Locally stored} & \yes\\\hline
 
 \ref{iot:iotframework} & IoT-framework & \yes & Server & Centralized
 & \good{Apache license 2.0} & \yes & \good{Locally stored} &
 \yes\\\hline
 
 \ref{iot:ifttt} & IFTTT & \maybe{Yes} & SaaS & Centralized & \no &
 \no & \good{No storage} & \maybe{Limited}\\\hline

 \ref{iot:kahvihub} & Kahvihub & \yes & Server & Centralized
 & \good{Apache license 2.0} & \yes & \good{Locally stored} &
 \yes\\\hline
 
 \ref{iot:linksmart} &  LinkSmart\textsuperscript{\texttrademark} &
 \maybe{Embedded} & P2P & Decentralized & \good{LGPLv3} & \no &
 \good{Locally stored} & \yes\\
 & & \maybe{devices} & & & \good & \bad & \good & \good\\\hline
  
 \ref{iot:myrobots} & MyRobots & \bad{Robots} & Robots PaaS &
 Cloud-based & \no & \yes & \maybe{2 levels} & \no\\\hline
 
 \ref{iot:niagara} & Niagara$^{AX}$ & \yes & M2M SaaS & Distributed &
 \no & n.a. & n.a. & n.a.\\\hline
 
 \ref{iot:nimbits} & Nimbits & \yes & Server & Centralized/ &
 \good{Apache license 2.0} & \yes & \good{3 levels} & \no\\
 & & \good{} & & Cloud-based & \good{} & \good{} & \good{} &
 \bad\\\hline

 \ref{iot:ninjablock} & NinjaPlatform & \maybe{Needs gateway} &
 PaaS & Cloud-based & \maybe{Open source hardware and Operating
 System} & \yes & OAuth2 & \no\\\hline

 \ref{iot:nodered} & Node-RED & \yes & Server & Centralized
 & \good{Apache license 2.0} & \no & \good{User-based} & \no\\
 & & \good & & & \good & \bad & \good{privileges} & \bad\\\hline

 \ref{iot:openiot} & OpenIoT & \yes & Hub & Decentralized &
 \good{LGPLv3} & \no & \good{User-based} & \yes\\
 & & \good & & & \good & \bad & \good{privileges} & \good\\\hline
 
 \ref{iot:openmtc} & OpenMTC & \yes & M2M client/ & Centralized/ & \no
 & \yes & Secured access & \no\\
 & & \good & Server & Cloud-based & \bad & \good & & \bad\\\hline

 \ref{iot:openremote} & OpenRemote & \maybe{Home devices} & Server &
 Centralized & \good{Affero GNU Public License} & \yes & \good{Locally
 stored} & \no\\\hline

 \ref{iot:opensense} & Open.Sen.se & \maybe{Ethernet enabled} &
 PaaS/SaaS & Cloud-based & \no & \yes & \maybe{2 levels} &
 \maybe{Limited}\\\hline

 \ref{iot:realtimeio} & realTime.io & \bad{Needs gateway} & PaaS &
 Cloud-based & \no & \yes & Secured access & \no\\\hline
 
 \ref{iot:sensorcloud} & SensorCloud\textsuperscript{\texttrademark} &
 \no & PaaS & Cloud-based & \no & \yes & n.a. & \no\\\hline
  
 \ref{iot:skyspark} & SkySpark & \no & SaaS & Centralized/ & \no &
 \yes & n.a. & \no\\
 &  & \bad{} &  & Cloud-based & \bad{} & \good{} & & \bad\\\hline
  
 \ref{iot:swarm} & Swarm & \yes & PaaS & Cloud-based & \maybe{Client
  is  open source (unknown license)} & \yes & n.a. & n.a.\\\hline
 
 \ref{iot:tempodb} & TempoDB & \no & PaaS & Cloud-based & \no & \yes &
 Secured access & \no\\\hline
  
 \ref{iot:terraswarm} & TerraSwarm & \yes & OS & Decentralized & n.a. & n.a.
 & n.a. & \yes\\\hline
  
 \ref{iot:thethingsystem} & The thing system & \maybe{Home devices} &
 Server & Centralized & \good{M.I.T.} & \yes & \good{User's choice} &
 \no\\\hline 
 
 \ref{iot:thingbroker} & Thing Broker & \yes & Server & Centralized &
 \yes & \yes & \good{Locally stored} & \no\\\hline
  
 \ref{iot:thingspeak} & ThingSpeak & \yes & Server  & Centralized/ &
 \good{GNU GPLv3} & \yes & \maybe{2 levels} & \maybe{Limited}\\
 & & \good{} & & Cloud-based & \good{} & \good & \maybe & \maybe{}\\\hline
  
 \ref{iot:thingsquare} & ThingSquare & \maybe{Embedded} &
 Mesh & Cloud-based & \maybe{Gateway firmware} & \yes & \no & \no\\
 & & \maybe{devices} & & & \maybe{is open source} & \good & \bad &
 \bad\\\hline 

 \ref{iot:thingworx} & ThingWorx & \yes & M2M PaaS & Cloud-based & \no
 & \yes & \good{User-based} & \yes\\
 & & \good & & & \bad & \good & \good{privileges} & \good\\\hline

 \ref{iot:wotkit} & WoTkit & \yes & PaaS & Cloud-based & \no & \yes &
 Secured access & \yes\\\hline
 
 \ref{iot:xively} & Xively & \yes & PaaS & Cloud-based &
 \maybe{Libraries are open source (BSD 3-clause), platform is not} &
 \yes & Secured access & \yes\\\hline
\end{tabular}
\end{table*}

Table~\ref{table:IoTplatforms} lists the surveyed platforms and
summarizes some characteristics which are seen by the authors as
fundamental for meeting the expectations of the users and application
developers.  
Hence, this table aims to provide quick visual information for those
interested in selecting the most appropriate IoT platform to be
deployed in their environment.
To improve the clarity of the table, we applied a color code to the
table cells. Specifically, the green color indicates that a particular
platform's characteristic fits the expectations of the users of the platforms, 
while the red color indicates a mismatch between the characteristic of the 
platforms and the expectations of the users. An intermediate orange
color has been added to indicate partial fitting.

In Table~\ref{table:IoTplatforms}, Column~\emph{a)} enumerates the
types of devices that are supported by the platform.  
Platforms that require a proprietary gateway to connect IoT devices
are dependent on the platform providers to respond to emerging
technologies, thus limiting the reactivity of the platform to adopt new
protocols and support an increasing number of heterogeneous IoT
devices.

Column~\emph{b)} describes the type of the IoT platform. In most
cases, the platforms are provisioned from a cloud, as shown in 
\figurename{~\ref{fig:cloud_platform}}, either in a  form of a 
Platform-as-a-Service (PaaS) or a Sotfware-as-a-Service (SaaS). 
The PaaS refers to the platforms that provide cloud
computing services for IoT devices and data. 
The services include, but are not restricted to storage facilities,
devices  management, device connectivity, backup mechanisms or online
support. 
By contrast, SaaS focuses on the mashup of data using cloud computing
capabilities. We added an additional Machine-to-Machine (M2M) tag if
the platform targets primarily this part of IoT~\cite{Kim2014}.

The type of architecture is shown in the Column~\emph{c)}. 
While the independent deployments are usually centrally controlled
(see \figurename{~\ref{fig:local_platform}}), the decentralized
deployments
(\emph{LinkSmart\textsuperscript{\texttrademark}}~\citeiot{iot:linksmart}
or \emph{OpenIoT}~\citeiot{iot:openiot}) include multiple sub-networks
of sensing and actuating devices (referred to as sites in 
\emph{LinkSmart\textsuperscript{\texttrademark}} and hubs in
\emph{OpenIoT}) that are independently controlled.

Note that no color code is used for columns \emph{b)} and \emph{c)} 
as we believe that different types of platforms and architectures are 
needed in different deployment environments.
For example, a decentralized PaaS, such as
\emph{H.A.T.}~\citeiot{iot:hat}, is ideal for a home environment
while a cloud-based solution like \emph{Xively}~\citeiot{iot:xively}
is more appropriate for a large deployment of sensors and actuators
(e.g., smart factory).

The table also includes information about the openness of the
platforms, the availability of a Representational State Transfer
(REST) API, as well as data access control and service discovery
mechanisms.

A number of open-source platforms are considered more
promising compared with the proprietary alternatives for the
following reasons. 
First, the use of the open source is expected to enable the faster
integration of new IoT solutions across the application
domains. Second, the use of the open source has been reported to speed
up the adoption of a software technology in a bottom-up
fashion. Finally, when seen from the social surplus perspective, the
industry based on the open-source platforms has been found to provide larger total welfare, compared with the industry structures based
on proprietary platforms~\cite{Economides2006}.

Only a few platforms do not have a REST API. This demonstrates that
the current IoT services will tend to become more and more like
traditional web services (i.e., Web of Things~\cite{Perez2014}). 
In particular, IoT service mashups and data
analytics will be key integrators for the future of IoT
technologies~\cite{Qin2011,Ma2013,Tsai2014}. We noted that only a few
platforms have integrated some type of service discovery mechanisms,
even in a very simplified fashion.
A comprehensive survey on discovery protocols for constrained M2M
communications can be found in~\cite{Villaverde2014}.

\subsection*{Security and privacy of IoT platforms}
One of the fundamental criteria for IoT platforms is the need to
include efficient and reliable privacy and security
mechanisms~\cite{Yan2014,Roman2011,Zhang2014,Roman2013,Satyadevan2015}.
In \cite{Satyadevan2015}, Satyadevan \textit{et al.} survey five IoT
platforms (including platforms \ref{iot:carriots},
\ref{iot:linksmart}, \ref{iot:thingspeak} and \ref{iot:xively} of our
survey) with respect to security and trust management.
The survey suggests that cloud-based IoT platforms are prone to
traditional web and network security attacks such as Denial of Service
(DoS), man-in-the-middle, eavesdropping, spoofing and controlling
attacks. 
A survey of low-level protocols for ensuring security and privacy in both
centralized and distributed IoT scenarios is presented in~\cite{Roman2013},
and the research community constantly aims to improve protocols to address these 
security challenges. An example is the work proposed by Asokan \textit{et
  al.}~\cite{Asokan2015} to secure large swarms of embedded devices
while overcoming the limitations of these constrained devices (i.e.,
memory, computation, communication, latency and energy consumption
constrains). 
A number of areas are critical for the widespread adoption of IoT but not yet fully 
addressed by IoT platforms; many of these have been listed and analyzed in the above surveys, including: 
(i) device authentication, (ii) communication and physical privacy, 
(iii) data storage protection, (iv) device protection, (v) trust
management, (vi) governance and (vii) fault tolerance.

In this article, we limit our analysis of the security and privacy issues to the protection
mechanisms for data storage and data access
available on the IoT platforms. Meanwhile, for
more comprehensive discussions on the other security, privacy and trust
challenges pertaining to IoT platforms, we invite the interested readers
to refer to \cite{Zhang2014,Roman2013,Satyadevan2015,Asokan2015}.

To authenticate users, most of the cloud-based platforms use the standard 
protocol OAuth~2.0~\cite{rfc6749} while
centralized servers required only a local access to the machine.
We evaluated the expectations in term of privacy as the flexibility of
the access control offered by the platform. 
Throughout our evaluation, we noted four types of access granularity
from the  basic private or public choice (i.e., 2-level for
\emph{MyRobots}~\citeiot{iot:myrobots} or
\emph{Open.Sen.se}~\citeiot{iot:opensense}) to a fine-grained access
control where the data could be either private, protected, public or
anonymous (i.e., 4-level for
\emph{EveryAware}~\citeiot{iot:everyaware}). 
In our opinion, the latter is the only one having the necessary
flexibility to maximize the re-usability of the data by remote
third-party services.

\section{Gap analysis}\label{sec:gap}

In the previous section, we presented the characteristics of IoT
platforms that are the most important to users and application
developers.
However, multiple gaps can be identified in the functionality offered
by these platforms.
Therefore, we present in this section a gap analysis, summarized in Table~\ref{tab:gapAnalysis},
that aims to
evaluate the maturity of the current solutions by assessing their
shortcomings along several dimensions. 
The dimensions covered by the analysis include (i) the extensibility of the
platform in terms of supporting heterogeneous sensing and actuating
technologies, (ii) the data ownership and its implications for
security and privacy, (iii) the data processing and sharing for
supporting new services, (iv) the support of application developers, 
and (v) the completeness of an IoT ecosystem. 
Then, we extend the gap analysis to dedicated IoT marketplaces that
(vi) support the deployment of IoT applications and services.

\subsection{Integration of sensing and actuating technologies}
\label{sec:gap_devices}

The essence of an IoT platform is to enable the secure connection of a
multitude of heterogeneous sensing and actuating devices, having
different constraints and capabilities, to the Internet. 
In the absence of de-facto communication standard(s), 
the sensing and actuating devices by different vendors may subscribe 
to different interaction patterns, and may implement different subsets 
of available communication protocols.
As a result, arguably, the value of an IoT platform grows proportionally 
with the number and the versatility of the supported devices. 
An ideal IoT platform would offer a
pool of standardized communication protocols where the device
manufacturer may select the appropriate protocols (e.g., CoAP for
constrained devices~\cite{Bormann2012}). 
In the case of passive devices (e.g., RFID-enabled) or constrained
devices, the connectivity relies on a mediating gateway (see 
\figurename{~\ref{fig:siloed_architectures}}) that must be fully
controlled by the platform user, alike the \emph{NinjaBlock}~\citeiot{iot:ninjablock},
which provides open-source hardware and firmware for the gateway.

For a smooth integration with sensing and actuating devices, 
it is essential that the IoT communities establish standardized
protocols for all devices, as it is currently done for highly
constrained devices by the IETF~\cite{Ishaq2013} or for M2M
communications by IEEE1888, ETSI M2M and
3GPP~\cite{Klinpratum2014}\footnote{ 
More details on standardization bodies and protocols can be found
in~\cite{Borgia2014}.}.
Presently, protocols for constrained devices are supported by
\emph{OpenRemote}~\citeiot{iot:openremote} (KNX, Z-Wave, X10, etc.), 
\emph{LinkSmart\textsuperscript{\texttrademark}} (ZigBee), 
\emph{ARM mbed}~\citeiot{iot:armmbed} (MQTT, CoAP) and \emph{ThingWorx}
(MQTT);
the others assume the use of
relatively powerful devices capable of supporting traditional web
protocols. It shall be noted that for some platforms, such as 
\emph{LinkSmart\textsuperscript{\texttrademark}}, the support for
constrained devices protocols is implied though the publicly
available documentation is insufficient for judging the extent of such
support.
Meanwhile, \sloppy{\emph{SensorCloud\textsuperscript{\texttrademark}}}
\citeiot{iot:sensorcloud}, \emph{SkySpark}~\citeiot{iot:skyspark} or
\emph{TempoDB}~\citeiot{iot:tempodb} require full-fledged HTTP
end-point to upload the data, assuming powerful devices capable of
supporting traditional web protocols and do not integrate device
communication protocols into their solutions.
Finally, \emph{IFTTT}~\citeiot{iot:ifttt}, which communicates with
both device manufacturers and web service providers,  ``adjusts'' to
the vendors' needs (e.g., to the needs of Belkin) to extend the
platform to new technologies. 
It shall be emphasized that there is no de-facto communication 
protocol suit, and this makes the task of interfacing heterogeneous 
devices more challenging and hence more expensive.
In addition, as previously mentioned in Section~\ref{sec:review}, 
IoT platforms do not integrate sufficient security and privacy protocols
to satisfy the integrity of the data and the management of connected 
devices~\cite{Roman2013,Satyadevan2015}.

The current IoT solutions address the issue of interfacing heterogeneous 
devices differently. Generally, the interoperability with devices is ensured 
either by implementing a gateway that can be expanded, e.g., with the 
help of plug-ins, to support new types of devices whenever needed, 
or by mandating the device vendors to use protocols from a limited set 
of supported ones. 
For example, the \emph{realTime.io}~\citeiot{iot:realtimeio} platform
proposed a connection of sensors via a proprietary gateway
(\emph{ioBridge} which even requires the use of a proprietary
transport protocol, \emph{ioDP}), while the
\emph{ThingWorx}~\citeiot{iot:thingworx} and
\emph{OpenMTC}~\citeiot{iot:openmtc} platforms use web sockets,
MQTT or other standard communication protocols to interconnect devices
to the platform. 
Some other platforms, such as 
\emph{the thing system}~\citeiot{iot:thethingsystem} or \emph{H.A.T.}
targets the  integration of devices present in ``smart homes'' and
``smart places'' environments. 
Other platforms, such as \emph{Fosstrack} \citeiot{iot:fosstrack},
only enable one type of technology which, in the case of
\emph{Fosstrack}, is RFID. 
Note however that either the heterogeneity of supported devices is limited, 
or the use of a gateway is necessary (Gap G1.1).
We believe that, in order to streamline the integration of new device types, 
standard object models for IoT devices, such as the models recommended in the recent
IPSO Smart Objects guidelines~\cite{IPSOAlliance2014} based
on the Lightweight M2M (LwM2M 1.0) specifications~\cite{OMA2013},
should be integrated widely by IoT platforms (Gap 1.2).
Furthermore, security mechanisms, such as in \cite{Asokan2015}, 
should be integrated to IoT platforms to provide secure management of 
IoT devices (Gap 1.3).

\subsection{Data ownership}
\label{sec:gap_ownership}

In our opinion, the enormous volume of data that would be generated by the
devices in the IoT mandates the data management to be at the core of IoT 
paradigm,  and it amplifies the need to maintain a certain degree of privacy 
and security~\cite{Roman2011}\footnote{ETSI Partnership Project oneM2M
has issued a  technical report for the
specification of security architecture for M2M
communications~\cite{oneM2M2015}. The standardization body has
classified the security solutions as (i) identification and
authentication, (ii) authorization and (iii) identity management.
In this study, we only considered the first two classes.}.
The owner of the data can thus be expected to have a full control over the placement 
of the data, as well as over who has the access rights to which portions of this data. 

Based on the information collected during our gap analysis of today's
IoT platforms, the data ownership has been a major concern for all the
platforms. 
For instance, the cloud-based platforms (e.g.,
\emph{Swarm}~\citeiot{iot:swarm}) ensure that the data collected and
stored by the platform remains the property of the customers. 
However, the full ownership of the data is rarely guaranteed.
In most cases, rather than storing and manipulating the 
data locally at the edge, the data is sent to the platform in a raw format, stored
unencrypted and very little information is presented on the security measures
taken to secure the data (Gap G2.1).

The majority of the listed platforms requires the use of access keys
or other access control mechanisms to get \emph{read} or \emph{write}
permissions.
The access rights are either determined by
the end users of the devices, through a web interface
(\emph{ThingSpeak}~\citeiot{iot:thingspeak},
\emph{Nimbits}~\citeiot{iot:nimbits}), or are left for the application
provider to define when implementing the applications
(\emph{OpenRemote}, \emph{Swarm},
\emph{LinkSmart\textsuperscript{\texttrademark}}, 
\emph{Thing Broker}~\citeiot{iot:thingbroker}). 
Furthermore, the \emph{EveryAware} platform provides access 
to public data feeds to anonymous users, who do not require access keys.
Such overly strict or too relaxed privacy settings do not provide enough 
control over the data (Gap G2.2).

Only solutions, where the data is stored locally (e.g., H.A.T. or OpenRemote), 
truly offer the full ownership of the data to the end-user. 
We suggest that future IoT solutions must have algorithms and mechanisms
for the data owner to give access only to a predefined set of the resources, 
and that the raw data must remain under control of the end-user. 
For instance, if the data owner is willing to archive data using a service
offered by a PaaS, he must be able to encrypt the data or process it before
sending it to the cloud.  
Further, since too strict or too relaxed privacy settings do not provide enough 
control over the data, in future IoT solutions, fine-grained data visibility must 
be coupled to local storage functionalities to re-attribute the ownership of 
the data to the users. 

\subsection{Data processing and data sharing}\label{sec:gap_sharing}

IoT data can be large in terms of volume and the applications typically 
have real-time requirements. 
IoT data streams are unbounded sequences of time-varying data elements. 
This data could often be unreliable, incomplete, and have different
qualities and out-of-order arrival problem, and communication
loss~\cite{Gubbi2013}. 
Furthermore, this data is represented in different
formats and various models. For example, it is a challenge to directly
utilize low-level data provided by sensors without a well-defined knowledge
model.

Data and knowledge behind data are the core of the wealth produced by the IoT.
Data processing and sharing mechanisms should be
developed to ensure that IoT data can be utilized in applications to its best.
Today's IoT solutions either do not support, or have limited support for the 
processing and sharing of data streams.
Yet, it remains possible to combine multiple streams into a single application 
if one knows the URI to the desired sources of information, but this represents
a technical challenge for application developers.
The \emph{Ericsson's IoT-Framework}~\citeiot{iot:iotframework}
provides mechanisms to integrate virtual streams (e.g., from 
external data sources) that can be combined with local streams for
visualization or statistical analysis and data predictions.
Moreover, different data processing techniques are adapted for IoT. 
For example, Tsai \textit{et al.}~\cite{Tsai2014} survey data mining
technologies for IoT. Su \textit{et al.}~\cite{Su2015} study how to
embed semantics on IoT devices and Maarala \textit{et
  al.}~\cite{Maarala2014} extend this research with processing large
IoT data in city traffic scenarios.
Nevertheless, the aggregation of these available data processing
techniques within IoT platforms is still limited (Gap G3.1).

The principle of data fusion has been addressed by the
\emph{Node-RED} tool~\citeiot{iot:nodered}, which enables the
composition of IoT data and devices with the concept of 
\emph{data flows}. In~\cite{Blackstock2014}, Blackstock and Lea
integrated the \emph{Node-RED} composer to the WoTKit
processor~\citeiot{iot:wotkit} to enable the creation of distributed
data flows.
Hence, such mechanisms support the creation of innovative and enriched
web-of-things contents. 
We suggest that such mechanisms should be integrated into IoT middleware
systems to perform similar operations on data streams. 
The current gap is in processing these streams efficiently and 
handling different formats and models (Gap G3.2). 
Here, efficient processing means (i) processing IoT data
considering computing, storage, communication, and energy
limitations of IoT environments; and (ii) the timely generation of useful
knowledge for IoT applications before it becomes outdated.
Meanwhile, to cope with big IoT data, most
IoT platforms shall have a high processing throughput. 

To mitigate the gap above, edge analytics solutions
(i.e., closer to where the data is being produced), 
such as \emph{cloudlets}~\cite{Lewis2014}, are now available for 
constrained deployments. We believe that future IoT platforms should include
\emph{cloudlets}-like technologies to enable local IoT networks to perform
edge analytics. 
Edge analytics contributes to maximize energy efficiency, reduce
privacy threats and minimize latencies. 
The \emph{Kahvihub} platform~\citeiot{iot:kahvihub} envisions to
support this for constrained devices, by providing sandboxed
execution platforms for IoT services. As a result, networks of
heterogeneous devices can collaboratively analyze the data that they
produce. 
Sandboxing IoT application has also been addressed in~\cite{Kovatsch2012}.

IoT devices produces low-level data, which is often
unreliable, incomplete, disordered, and even lost. 
Therefore, fault management is essential for IoT platforms.
Availability of input data streams for IoT platforms is often undetermined. 
Hence, additional challenges are introduced to guarantee the  complete data processing result (Gap G3.4). Moreover, intrusion detection, prevention, and recovery mechanisms should be developed in IoT platforms, which will help IoT entities to protect their data and services ~\cite{Roman2013}.

Finally, in order to find the relevant data streams that are available, these
streams should be listed in dedicated data catalogs where context information may be
used to provide efficient discovery mechanisms. 
Semantic indexing can be used on these catalogs and other metadata available on the
IoT devices~\cite{He2012}. The efficient processing of IoT data
from multiple external sources is still an open issue.
From all the platforms reviewed in this article, only four (i.e.,
\emph{IoT-Framework}, \emph{Kahvihub}, \emph{ThingWorx} and
\emph{Xively}) integrated a search mechanism for data streams. 
In the case of \emph{IoT-Framework}, the search mechanism
is performed through geolocalization, tagging or data types.
However, the search was limited to the streams available on the platform, 
whereas the search through multiple platforms is hitherto unavailable (gap G3.3). 
Recent research efforts have been invested towards this direction with
\emph{HyperCat}, a lightweight JSON-based URI catalog that references services
provided by IoT platforms~\cite{Lea2014a}.

\begin{table*}[!ht]
 \caption{Summary of the gap analysis}\label{tab:gapAnalysis}
 \centering
 \scriptsize
 \begin{tabular}{|m{.09\textwidth}|m{.13\textwidth}|m{.16\textwidth}|
     m{.175\textwidth}|m{.15\textwidth}|m{.14\textwidth}|}
  \hline
  {\small Category} & {\small Current status} & {\small Expectations} &
{\small Gaps} & {\small Problems} & {\small Recommendations}\\
  \hline\hline
  Support of heterogeneous devices & 
  Platforms assume smart objects to talk HTTP or require gateway & 
  {\tiny$\bullet$}Devices must be easily and securely integrable to the IoT platform without a
  gateway 
  \newline{}{\tiny$\bullet$} Unified resources and simplify usability &
  \textbf{G1.1} Support of constrained devices
  \newline{}\textbf{G1.2} Standardized IoT devices models
  \newline{}\textbf{G1.3} Secure authentication, identification of management of IoT devices &
  {\tiny$\bullet$} Heterogeneous interactions
  \newline{}{\tiny$\bullet$} Protocol standardization
  & 
  {\tiny$\bullet$} Relying on standard protocols (e.g., CoAP, LwM2M, MQTT)
  \newline{}{\tiny$\bullet$} Integration of state-of-the-art security and privacy protocols\\
  \hline
  Data\newline{}ownership & 
  Mainly given to the end-user but with very simple privacy policies & 
  {\tiny$\bullet$} Full control given to the owner of the data
  \newline{}{\tiny$\bullet$} Local storage
  \newline{}{\tiny$\bullet$} Fine-grained data visibility model & 
  \textbf{G2.1} Manipulation of data in edge devices
  \newline{}\textbf{G2.2} Self-storage &
  {\tiny$\bullet$} Security of the data storage
  \newline{}{\tiny$\bullet$} Device constrains to store data and provide
  secure access control & 
  Algorithms and mechanisms available to the data owner to limit the access only to a predefined set of the resources\\
  \hline
  Data processing \& sharing &
  {\tiny$\bullet$} Nonuniform data sharing format
  \newline{}{\tiny$\bullet$} Sharing is performed via nonuniform REST API & 
  {\tiny$\bullet$} Uniform data format across multiple platforms.
  \newline{}{\tiny$\bullet$} Pub/Sub mechanism and data catalogs 
  \newline{}{\tiny$\bullet$} Edge analytics &
  \textbf{G3.1} Data processing is not well integrated in IoT platforms
  \newline{}\textbf{G3.2} Efficient processing for data formats and models
  \newline{}\textbf{G3.3} Data analytics is only available in cloud-based solutions 
  \newline{}\textbf{G3.4} Data catalogs are missing &
  {\tiny$\bullet$} Complex identification system to access data
  \newline{}{\tiny$\bullet$} Fusion efficiently data streams from multiple data 
  catalogs
  \newline{}{\tiny$\bullet$} IoT devices have limited computing capabilities & 
  {\tiny$\bullet$} Data catalogs with semantic indexes
  \newline{}{\tiny$\bullet$} Uniform and interoperable data
  models
  \newline{}{\tiny$\bullet$} Integration of data processing technologies in platforms
  \newline{}{\tiny$\bullet$} Cloudlet-like solution for edge analytics
  \\\hline
  Developer support & 
  {\tiny$\bullet$} REST API to access the data or devices handled by the
  platform 
  \newline{}{\tiny$\bullet$} Applications are for internal use rather than for
  sharing (except IFTTT) & 
  {\tiny$\bullet$} Use of a common API to ease the development of cross-platform
  applications
  \newline{}{\tiny$\bullet$} Domain Specific Language (DSL) dedicated to
  cross-platform application development &
  \textbf{G4.1} Application mash-up APIs
  \newline{}\textbf{G4.2} Limited presence of SDKs 
  \newline{}\textbf{G4.3} Absence of DSL with higher abstraction level primitives &
  {\tiny$\bullet$} Require standardization of application interactions dedicated
  to the IoT
  \newline{}{\tiny$\bullet$} IoT app store are missing &
  IoT platforms must provide SDKs and APIs that maximize the re-usability of the
  services provided by their platform\\
  \hline
  Ecosystem formation &
  Platforms provide useful building blocks, storage and run-time environment for
  application developers &
  {\tiny$\bullet$} Platform easily expandable by the developers and offering
  them incentives to contribute
  \newline{}{\tiny$\bullet$} Cross-platform sharing of applications and
  services
  \newline{}{\tiny$\bullet$} Local composition of services&
  \textbf{G5.1} Low platform expandability
  \newline{}\textbf{G5.2} Limited monetizing possibilities
  \newline{}\textbf{G5.3} Limited support for cross-platform integration &
  {\tiny$\bullet$} Silos of platform-specific solutions
  \newline{}{\tiny$\bullet$} User's using multiple platforms may not be able to
  aggregate the whole data into a single application& 
  {\tiny$\bullet$} Financial incentives for developers shall be offered
  \newline{}{\tiny$\bullet$} A broker is needed to ease cross-platform
  integration
  \newline{}{\tiny$\bullet$} Models to contextually define IoT applications to
  simplify their discovery by the end-users\\
  \hline
  IoT\newline{}marketplace &
  {\tiny$\bullet$} Limited applications sharing
  \newline{}{\tiny$\bullet$} Limited (usage-based) charging of the end
  users of these applications &
  {\tiny$\bullet$} Dedicated IoT data catalogs, IoT app store and IoT device
  store
  \newline{}{\tiny$\bullet$} Ability to advertise, deliver and charge for the 
  use of applications and data
  \newline{}{\tiny$\bullet$} Validate applications against policies &
  \textbf{G6.1} Application, data and device catalogs dedicated to the IoT
  are generally missing
  \newline{}\textbf{G6.2} The billing (based on fixed fees, usage, or other
  metrics) of the end-users of the data is generally missing &
  An ecosystem of independent application developers, device manufacturers, and
  end-users all supporting the platform is needed for the demand for marketplace
  to appear and sustain &
  The marketplace functionality shall be provided by future IoT platforms\\
  \hline
 \end{tabular}
\end{table*}

\subsection{Support of application developers}
\label{sec:gap_dev}

In order to foster an expedited development of applications, 
the IoT platforms are expected to provide the developers with streamlined 
application programming interfaces (APIs) to their functionality, 
preferably with the help of higher abstraction level primitives. 
Further, to enable an efficient development of cross-IoT-platform applications, 
these APIs shall be uniform across the platforms, to the extent possible.

Today's IoT platforms almost all provide a public API to access the services.
The APIs are usually based on RESTful principles, 
and allow common operations such as \texttt{PUT}, \texttt{GET}, \texttt{PUSH} or \texttt{DELETE}. 
These operations support the interaction with the connected devices on the
platform, as well as the management of these devices. 
Only four of the studied platforms did not include a REST API for easing the
development of web services (i.e. \emph{Fosstrack},
\emph{LinkSmart\textsuperscript{\texttrademark}},
\emph{IFTTT} and \emph{OpenIoT}), but use different interaction means.
Nonetheless, the other platforms uses nonuniform\footnote{Nonuniform
  in this context means that every platform provides custom APIs and
  data models as standards such as HyperCat~\cite{Burt2014} are not
  yet widely adopted.} REST APIs and data models which complicates the
mashing up of data across multiple platforms (Gap 4.1; see 
Section~\ref{sec:gap_sharing}).

Many platforms also offer libraries, which are in some cases open-source (e.g.
\emph{AirVantage\textsuperscript{\texttrademark}}~\citeiot{iot:airvantage},
\emph{Exosite}~\citeiot{iot:exosite}, \emph{IoT-Framework} or
\emph{Xively}), that are bindings for different programming languages
to the REST API available on the platforms.
However, these bindings libraries do not greatly improve the support to
application developers in using the services provided by the platforms as they
only include basic functionalities, e.g., connection to the platform with access
keys (Gap 4.2).
To some extent, some platforms such as \emph{ThingSpeak} enable the creation of 
widgets written in Javascript, HTML and CSS that may be distributed on the
platform to other users.
Alternatively, the \emph{Carriots\textsuperscript{\textregistered}}
platform~\citeiot{iot:carriots} provides a full
Software Development Kit (SDK) written in Groovy for application developers. 
We believe that this approach should be more generalized within IoT solutions to
maximize usability of the services provided by the IoT platforms.

In addition to APIs, a Domain Specific Language (DSL) could be defined 
to simplify the development of IoT applications, also by offering functional 
primitives describing the problem and solution space at a higher abstraction level. 
For instance, primitives for querying the data stream catalogs, fusing and 
aggregating data should be available to the developers in order to speed up the
process of developing cross-platform data-centric applications; such DSL, however,
are largely non-existent at the time of writing (Gap 4.3).

\subsection{Toward IoT ecosystem formation}
\label{sec:gap_ecosystem}

The success of an IoT platform is dependent on the existence of a business
ecosystem of firms where the buyers, suppliers and makers of related products or
services, as well as their socio-economic environment, collectively provide a 
variety of applications, products, and services to the end-users of IoT~\cite{Mazhelis2012}. 
By offering a common set of assets, that are shared by the ecosystem members and
are essential for their products and services, such platform shapes a core of
its ecosystem. 

To prosper, the platform, besides performing an essential IoT function or
solving an essential IoT problem, should be easily expandable by the developers
of the complementing products or applications based on it, and should provide
them with incentives to innovate and contribute to the platform
\cite{Gawer2008}. In other words, the platform shall attract the developers of
add-ons and applications, thereby enabling a bottom-up formation of the ecosystem around it.

Today's IoT platforms claim to solve some of the essential problems of
application developers, and are generally open for third-party application
creators. However, only open-source platforms can be expanded rapidly to cope
with the emergence of new technologies. Proprietary platforms do not allow to
add reusable components or add-ons  to the platform, except recipes in 
\emph{IFTTT} and  third-party tools integration for 
\emph{ThingWorx} (Gap 5.1), and monetizing possibilities for platform complementers 
are absent or limited to integration services, e.g., \emph{OpenRemote} (Gap 5.2).

In order to allow to treat the IoT domains as a single converging
ecosystem that provides innovative products and services and permits
an economy of scale, an IoT platform broker is needed. 
Such a broker will facilitate the sharing of applications and services
across space and time, and across platform-specific IoT sub-ecosystems.
However, the possibility of multi-platform brokerage has not been
investigated in depth and the resulting IoT ecosystem represents a
multitude of fragmented IoT vertical silos (Gap 5.3).

Still, this vision of new IoT ecosystem formation is shared by the
Terra Swarm Research Center for the
\emph{TerraSwarm}~\citeiot{iot:terraswarm} and by the Technology Strategy
Board\footnote{\url{https://www.innovateuk.org/}} 
with the specification of \emph{HyperCat} to solve interoperability issues 
among IoT solutions.

\subsection{Dedicated IoT marketplaces}\label{sec:marketplace}
\label{sec:gap_marketplace}

\begin{figure}[t]
 \centering
 \includegraphics[width=\linewidth]{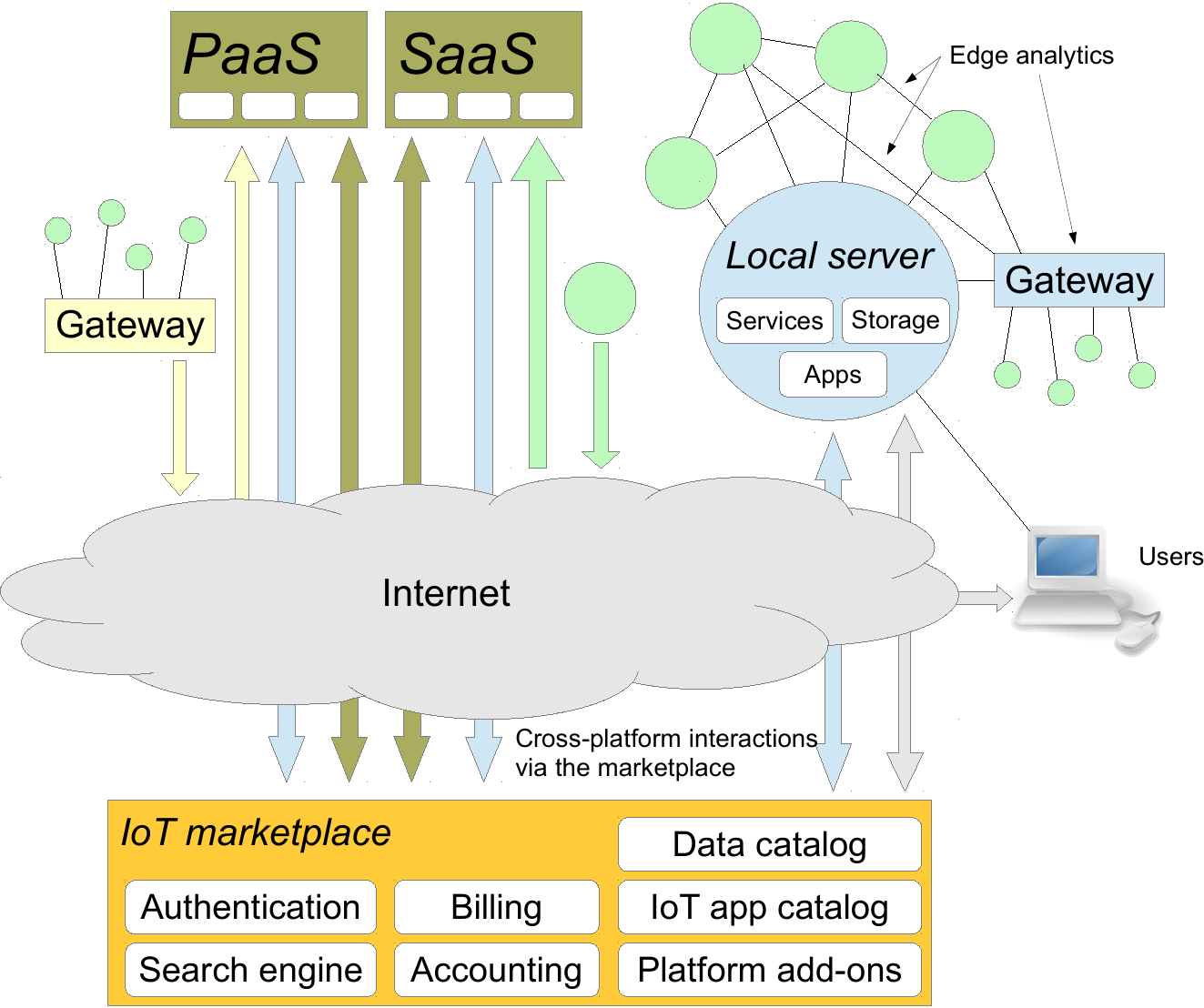}
 \caption{
 	\textbf{End-to-end interactions between users, smart devices, the platforms via the marketplace.}
 	\textsl{The IoT marketplace allows cross-platform interactions and drives the development of new 
    business opportunities (e.g., billing of IoT data). White rectangles within platforms represents 
    functional elements (e.g., search engine or apps) but text was omitted in cloud-based architecture to improve visibility. 
    Check \figurename~\ref{fig:cloud_platform} for more details.}
 }
 \label{fig:market_interactions}
\end{figure}

Software application marketplaces are aimed at facilitating the discovery,
purchase, and distribution of the applications. 
These marketplaces can be exemplified with hardware-specific and
centrally-controlled solutions, such as \emph{Apple App Store} or \emph{Google
Play}, or hardware-agnostic marketplaces, such as \emph{Good}, \emph{Handster},
\emph{Nexva}, and \emph{SlideMe}. 
The availability of such marketplaces is crucial for the dissemination of
software innovations in general, and IoT innovations in particular
\cite{Kortuem2010}.  

These marketplaces address the needs of the application providers and
users, and alternatively, the needs of the platform vendors and
platform operators.
However, the traditional application stores are seemed to have limitations as far as
IoT applications are concerned. 
Namely, to the best of our knowledge, none of the contemporary application
stores support the delivery of purchased software to the connected devices other
than the mobile terminals (e.g. smartphones and tablets) supported by the
platform (Gap 6.1). 
Among IoT platforms, some platforms have dedicated application stores (e.g. 
ThingWorx) but only some (\emph{IFTTT}) allow the applications to be
publicly shared, and only some (\emph{OpenIoT}) promise to enable the
(usage-based) charging of the end users of these applications (Gap 6.2).
Moreover, one of the key challenges of IoT is to exploit all the data that is currently
being produced by businesses.
According to McKinsey\footnote{\url{http://www.mckinsey.com/insights/business_technology/the_internet_of_things_the_value_of_digitizing_the_physical_world}},
businesses already collect tremendous volume of sensor data
but the data is only used for anomaly detection and control. 
However, data should also be used for optimization and prediction which provide the greater value, 
but businesses may lack the expertise to analyze and process their data.
This justifies the need for the development of new marketplaces for IoT
data that will thrive new business interactions (i.e., business-to-business).

The \emph{Windows Azure Data Market}\footnote{
\url{http://datamarket.azure.com/}} platform provides an example of a successful 
business model that could emerge from IoT data.
For instance, the platform allows businesses to publish data streams to the 
platform in order to make them available to a large number of application 
developers.
The platform offers the possibility of charging for the data consumption either 
by a time-defined subscription or by the amount of data to be consumed. 
The platform also allows publishing data streams free of charge.
The catalogs of data sources published on the platform is also browsable. 
We believe that the development of IoT dedicated platforms, similar to the
\emph{Windows Azure Data Market}, is a requisite to the sustainability of IoT
solutions. 
Figure~\ref{fig:market_interactions} illustrates the opportunities that emerge from the availability of dedicated IoT marketplaces.
Unlike in Figure~\ref{fig:siloed_architectures}, where interactions between IoT platforms is limited and difficult, the IoT marketplace allows the flow of IoT data across platforms.
Marketplaces should include authentication, billing, accounting,
as well as catalogs for IoT data and applications. 
Marketplaces could also be extended with an additional catalog for communication 
protocols (platform-specific) and for IoT devices/components to provide a complete 
solution for the IoT users.

\section{The perspective of the national Finnish IoT program}
\label{sec:survey}

In this section, we present the results of a survey conducted
among the partners of the Finnish IoT program~\cite{Tarkoma2013} 
on the importance of various key points for
the future development of IoT platforms, including the IoT dedicated
market places. Table~\ref{tab:organisations} lists the number of
survey participants.

\begin{table}[h]
  \centering
  \caption{Organization distribution}
  \label{tab:organisations}
  \begin{tabular}{|l|l|l|}
    \hline
    Type & Count & Percentage\\\hline
    Academia & 19 & 54.29\%\\
    SME & 7 & 20\%\\
    Large company & 9 & 25.71\%\\\hline
    Total & 35 & 100\%\\\hline
  \end{tabular}
\end{table}

\begin{figure}[t]
  \centering
  \includegraphics[width=\linewidth]{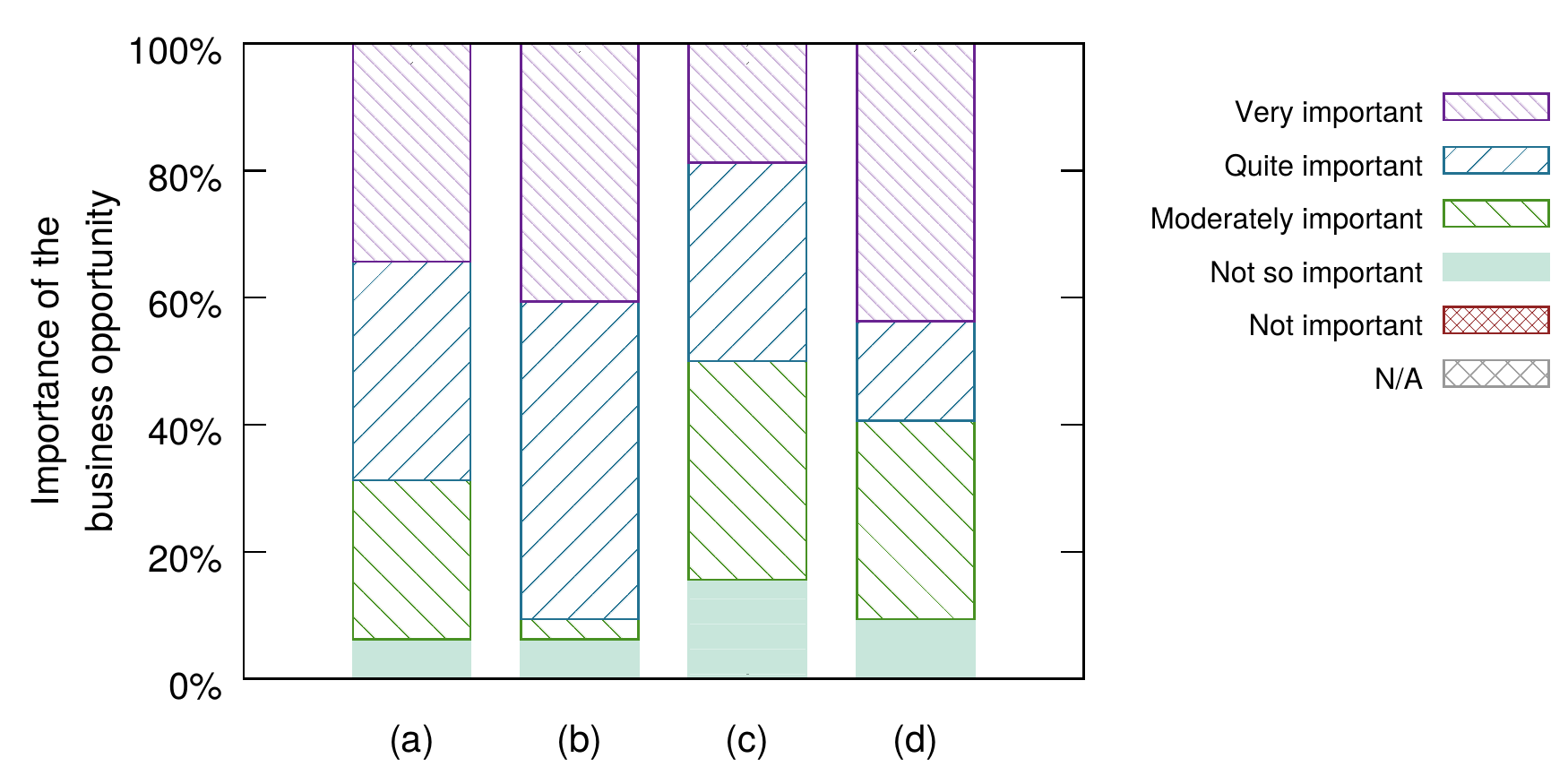}
  \caption{\textbf{Importance of the business opportunities} \textsl{for (a) sharing and
    selling data and/or applications in a controlled manner, (b)
    Maximizing re-usability of data to increase profit, (c) Searching
    for data/applications in an ad-hoc fashion and (d) reducing
    transaction costs of data/application acquisition.}}
  \label{fig:survey_business_opp}
\end{figure}

\figurename{~\ref{fig:survey_business_opp}} summarizes the results of
the survey regarding the possible business opportunities that could
emerge from filling out the gaps presented in the previous section.
As can be seen from the figure, survey respondents have designated the business opportunity of
maximizing re-usability of data as the most important (at 40\% very
important, as well as 50\% quite important). 
On the other hand, searching for data or applications in an ad-hoc
fashion raised less interest as less than 50\% of considered it quite/very important, 
and since 15\% of the project's experts declared
it of little importance.
Finally, the sharing and selling of data/applications as well as
reducing the cost of data/applications acquisition have raised
moderate interests. 

\begin{figure}[t]
  \centering
  \includegraphics[width=\linewidth]{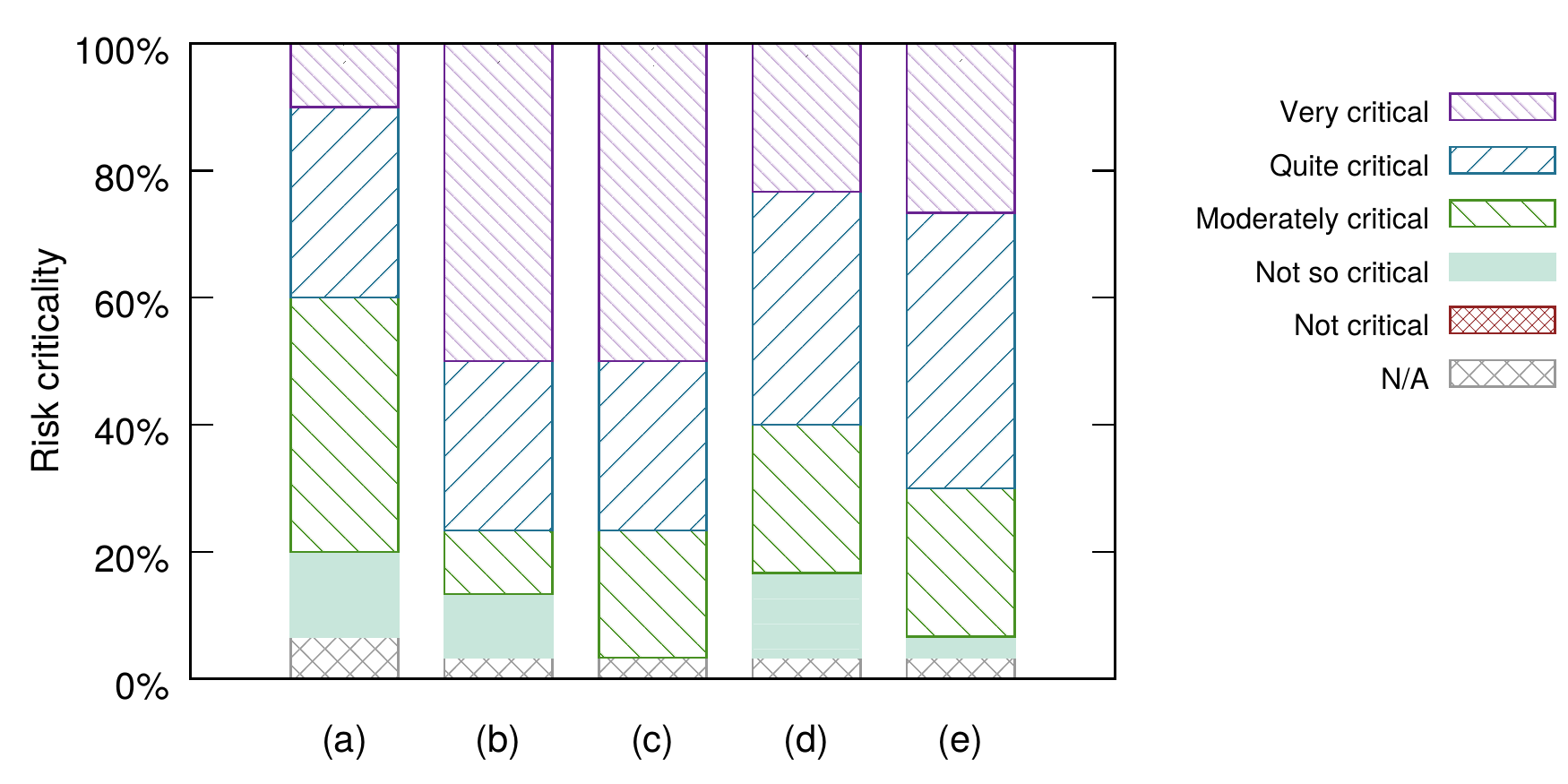}
  \caption{\textbf{Risk criticality} \textsl{of (a) direct sells preferred, (b) lack of
  data suppliers and application providers, (c) small customer base,
  (d) challenge of making generic applications and (e) fragmentation
  of the IoT landscape}.}
  \label{fig:survey_risk}
\end{figure}

We also asked our experts to evaluate the risks that may emerge
from developing the next generation of IoT platforms. 
As shown in \figurename{~\ref{fig:survey_risk}}, the most critical
risks are i) the lack of suppliers and application providers as well as
ii) having a too small customer base. 
In fact, these two risks are going hand in hand as a large customer
base attracts application developers and data suppliers, while the
latter attracts more customers.
Noteworthy, possible negative impact of the
introduction of IoT marketplace onto the traditional way of selling
directly has not been
defined as critical risk by our panel of experts (see \figurename{~\ref{fig:survey_risk}}(a)).
However, the risks coming out from the current verticality of the IoT
landscape have been found moderately critical, thus showing the
readiness of the IoT for more horizontal interactions between IoT
solutions (see \figurename{~\ref{fig:survey_risk}}(d) and \figurename{~\ref{fig:survey_risk}}(e)).

In the final stage of the survey, we asked our experts to evaluate the
most important features that must be integrated to IoT platforms with
regard to the gaps underlined in Section~\ref{sec:gap}. The features
are grouped by four different viewpoints; (i) application provider
viewpoint, (ii) data publisher viewpoint, (iii) platform provider
viewpoint and lastly (iv) the customer viewpoint.
The results of this evaluation has produced the following list of
features in a descending order of importance:
\begin{enumerate}
  \item Publishing applications: register and upload the
    applications, make applications discoverable and available for
    external parties (Application provider viewpoint).
  \item Available description or detailed information about the
    application or the data on the marketplace (Customer viewpoint).
  \item Purchasing the right to use the application or the data
    (Customer viewpoint).
  \item Publishing data: make the data discoverable and
    available for external parties through predefined interfaces (Data
    publisher viewpoint).
  \item Gathering information about resources usage by customers, as
    well as summarizing it into accounting records, e.g., for the
    purpose of charging and billing (Platform provider viewpoint).
  \item Setting or modifying the access rights separately to different
    views or portions of the data in order to maximize re-usability
    (Data publisher viewpoint).
  \item Gathering information about the sells and downloads of the
    applications (Application provider viewpoint).
  \item Searching for the applications based on type, payment details,
    rating (Customer viewpoint).
  \item Registering, unregistering, uploading and validating an
    application (Platform provider viewpoint).
  \item Managing platform subscriptions of customers (create, read,
    update, delete) (Platform provider viewpoint).
\end{enumerate}

This listing shows that eight out of the ten  most important features
selected by our experts are related to the IoT marketplace and generate cross-platform interaction as depicted in Figure~\ref{fig:market_interactions}, thus
comforting our view on the necessity of developing this type of
platform. The sixth most important feature is, on the other hand,
related to increasing the re-usability of the data by setting multiple
role-based views on the data or on selected portions of the data
(e.g., for only a time period of 24 hours).

\section{Recommendations for the development of IoT middleware}
\label{sec:recommendations}

In the previous sections, we evaluated the current IoT platform landscape with a
thorough gap analysis, that is summarized in
Table~\ref{tab:gapAnalysis}, and complemented the gap analysis with a
survey conducted among the experts of the national Finnish IoT program.
As a result, numerous gaps have been identified; 
furthermore, several recommendations were made in section~\ref{sec:gap} 
for the IoT platform vendors to expand their offerings so as to address these gaps. 
These recommendations included, among others, 
\begin{itemize}
\item leaning on standardized communication protocols to interface heterogeneous devices (Subsection~\ref{sec:gap_devices}),
\item adding the provisions for handling and processing data locally (Subsection~\ref{sec:gap_ownership}),
\item adding uniform data models, data catalogs, and the edge analytics capabilities (Subsection~\ref{sec:gap_sharing}),
\item offering streamlined APIs (Subsection~\ref{sec:gap_dev}), 
\item introducing cross-platform brokers and financial incentives for ecosystem players (Subsection~\ref{sec:gap_ecosystem}), and 
\item developing dedicated IoT makerplace(s) (Subsection~\ref{sec:gap_marketplace}). 
\end{itemize}

In this section, we return to these recommendations and complement them with further recommendations 
both concerning the short-term (easier to implement) and long-term (harder to implement) perspectives.
For the reader's convenience, these recommendations are shown in the right-most column of 
Table~\ref{tab:gapAnalysis}.

In the short-term perspective, the development of a basic IoT marketplace, as shown in
\figurename{~\ref{fig:market_interactions}},
serves as a repository for data streams and applications, should boost
tremendously the ability of the IoT landscape to fill partially in some of the 
gaps.
For instance, the immediate benefits would be:
\begin{itemize}
 \item \textbf{Data processing \& sharing}: the ability to request numerous external
 data streams to enrich local content. It would also enable users to publish some
 of their streams to third-parties;
 \item \textbf{Developer support}: the possibility for application developers to
 publish their products and reach a wide range of customers;
 \item \textbf{Ecosystem formation}: the increasing awareness about
   new innovations and possibility of creating new business models;
 \item \textbf{Market \& billing}: the ability to market/search for data and
   applications and sell/purchase the rights to use them.
\end{itemize}

From the viewpoint of middleware solutions, fine-grained access control must be 
implemented first to re-provision the user with the full ownership of his data.
Finally, SDKs should be provided to application developers to facilitate the creation of
the applications based on the platform.

In the long-term perspective, the marketplace would drive the uniformity for the 
REST APIs and the data models. It would also contribute to the standardization of
popular communication protocols as IoT device manufacturers will be encouraged
to comply to these open standards (e.g., ETSI, IETF, etc.) in order to
improve their visibility on the marketplace. 
Accounting functionalities must be implemented next to
strengthen ecosystems and permit a large scale economy.
Additionally, efficient search engines for data streams must be
developed to maximize the quality of services of IoT applications.
From the viewpoint of the middleware solutions, the development of a 
cross-platform DSL would provide massive support to application 
developers. Moreover, performing edge analytics (see \figurename{~\ref{fig:market_interactions}})
would help reduce the latency, the volume of data transported across the network and 
reduce threats on privacy and security (e.g., raw and risk-critical data may be pre-analyzed locally).

As a result, the marketplace plays a central role in connecting IoT actors and thus, allowing
cross-platform interactions for the IoT (different interactions are represented with separated colors in \figurename{~\ref{fig:market_interactions}}), and consequently creating more opportunities for data exchanges and
business operations. The marketplace will also allow the distribution of IoT-specific applications to a large number 
of IoT users as we currently experience with smartphone application stores, and thrive the economical growth of IoT which is expected to reach as much as \$19 billions (Cisco's forecast for 2020\footnote{\url{https://agenda.weforum.org/2014/01/are-you-ready-for-the-internet-of-everything/}}).

\section{Conclusions}\label{sec:conclusion}

In this article, we have evaluated a number of available IoT platforms, both
proprietary and open-source, that together form a representative sample of the IoT
platform landscape.
The IoT platforms were evaluated via a gap analysis that
outlined their capability to (i) support the integration of heterogeneous
hardware, (ii) provide sufficient data management mechanisms, (iii) support
application developers, (iv) support the formation of ecosystems, as well as (v) 
provide the dedicated marketplaces for the IoT. 
Collectively, these capabilities reflect the
needs of different players of the emerging IoT ecosystem,
including the device vendors, the application developers, the providers of
platforms and related services, and the end-users.

We complemented the gap analysis with a survey conducted among the
experts of the Finnish IoT program to
evaluate the business opportunities, risks and the most important features
that may emerge from filling in the highlighted gaps. 
Based on the results of the gap analysis and the survey, we compiled a
list of recommendations, both for short and long term perspectives. 
Our recommendations are aimed at filling in the identified gaps in 
contemporary IoT platforms and include, among others, the development of a 
dedicated IoT marketplace, the availability of SDKs and open APIs,
and the possibility to analyze data locally, flexibly control
access to the platform and its data, as well as providing data processing and sharing mechanisms.

\appendix
\section{Reviewed IoT Platforms}
\label{appendix}

\iotplatform{AirVantage\textsuperscript{\texttrademark}}
{https://airvantage.net/}\label{iot:airvantage} 
AirVantage\textsuperscript{\texttrademark} is a proprietary
cloud-based M2M dedicated platform that provides end-to-end solutions
to connect wireless-enabled devices to their platform. 
From an user viewpoint, the platform proposes interactive dashboards
for device management, and big data storage. The platform uses
open-source M2M dedicated development tools such as the framework 
\emph{m2m.eclipse.org}\footnote{\url{http://m2m.eclipse.org}}.
The platform also integrates the standard protocol MQTT.

\iotplatform{Arkessa}{http://www.arkessa.com/}\label{iot:arkessa}
Arkessa is a proprietary cloud-based M2M management architecture and
IoT platform. 
It includes the MOSAIC platform that enables devices to be easily
connected to many applications. Privacy with third-party applications
is done in similar way than Facebook or Linkedin. Ownership of the
data remains to the end-user. Arkessa provides an ecosystem of devices
and applications giving high flexibility to the end-user.

\iotplatform{ARM mbed}{https://mbed.org/}\label{iot:armmbed}
ARM mbed\textsuperscript{\textregistered} provides a device server,
that is proprietary, to connect constrained devices to the IoT. 
The platform proposes security solutions for embedded devices, such as
embedded Transport Layer Security (TLS).
It uses CoAP and RESTful API for creating M2M networks of constrained
devices.

\iotplatform{Carriots\textsuperscript{\textregistered}}
{https://www.carriots.com/}\label{iot:carriots}
Carriots\textsuperscript{\textregistered} is a proprietary cloud-based
platform (PaaS). REST API and Groovy SDK are available for web
application development. Data format supported are JSON and XML. 
The data is stored on the platform and access keys are required to
access it.

\iotplatform{DeviceCloud}{http://www.etherios.com/products/devicecloud/}
\label{iot:devicecloud}
DeviceCloud is a proprietary and cloud-based device management
platform (PaaS). The platform provides access the devices connected to
the platform via a REST API.

\iotplatform{Devicehub.net}{http://www.devicehub.net/}
\label{iot:devicehub}
Devicehub.net is a proprietary cloud-based platform which does not
provide a true REST API (using GET method to PUT data). Currently, the
documentation of the platform is too limited to provide more
information.

\iotplatform{EveryAware}{http://www.everyaware.eu/}
\label{iot:everyaware}
The EveryAware platform~\cite{Becker2013} provides an extendable data
concept that could be use to enhance the possibilities of sharing and
processing data feeds. The platform is running on a centralized
server. This platform was the one providing the finer-granularity of
data visibility with four different levels (details, statistics,
anonymous, none). A REST API has been integrated to access the data
(extendable data models).

\iotplatform{EveryWare Device Cloud\textsuperscript\texttrademark}
{http://www.eurotech.com/en/products/software+services/everyware+device+cloud}
\label{iot:everyware}
EveryWare Device Cloud\textsuperscript\texttrademark is a proprietary
cloud-based platform (PaaS) using a pay-as-you-go business model. 
A RESTful API supporting JSON and XML data formats, is integrated for
communication with the devices.
The sensors required to be connected to Eurotech gateway to be
connected to the cloud. 
A variety of applications and tools is available within the platform
to provide full end-to-end solution.

\iotplatform{EvryThng}{http://www.evrythng.com/}\label{iot:evrythng}
EvryThng is a proprietary centralized platform  (SaaS) that provides a
persistent presence on the Web of identifiable objects (RFID, NFC,
connected objects, etc.). It allows via RESTful API to store and
retrieve metadata as well as real-time data for these objects. The API
allows fine-access grained control to easy sharing of products
information. No search tools are available to find data
feeds. Billing is done on-demand. The EvryThng platform includes
standard protocols MQTT and CoAP.

\iotplatform{Exosite}{http://exosite.com/}\label{iot:exosite}
Proprietary cloud-based solution (PaaS) enabling vertical markets
(from devices to IoT solution).
Libraries for binding of the REST API with the Exosite platform are
open-source, available under the BSD license. 

\iotplatform{Fosstrack}{https://code.google.com/p/fosstrak/}
\label{iot:fosstrack}
Fosstrack is a closed-source SaaS platform to handle RFID
devices. Electronic Product Code (EPC) cloud have been developed on
top of the Fosstrack for fast deployments of RFID systems. Fosstrack
shows that the fragmentation of the IoT landscape is high. However,
the users stores RFID data on their own database accessed via a Tomcat
server.

\iotplatform{GroveStreams}{https://grovestreams.com/}
\label{iot:grovestreams}
GroveStreams proprietary cloud-based solution for analytics of data
from multiple sources. It uses a REST API and JSON data format. 
GroveStreams is an open platform, in the cloud, that any organization,
user or device can take advantage of. GroveStreams is free for small
users. Large users will only be billed for what they use.

\iotplatform{Hub of All Things}{http://hubofallthings.wordpress.com/}
\label{iot:hat}
The Hub-of-All-Things (H.A.T.) platform has as primary objective the
creation of multi-sided market platform to generate new economic and
business opportunities using IoT data generated by a ``smart home''. 
An important feature of the H.A.T. is that the data belongs to the
individual.
It enables the end-users to get control of their data, and
thus maintaining their expectations about privacy and other issues. 
In particular, the H.A.T architecture defines different kind of
applications (in-apps and out-apps). The ``in-apps'' (owned by either
residents, landlords or building managers) have their content enriched
by local data available on the private H.A.T, while ``out-apps'' may
be used by external platforms. 

\iotplatform{Ericsson IoT-Framework}
{https://github.com/EricssonResearch/iot-framework-engine}
\label{iot:iotframework}
The Ericsson IoT-Framework is a PaaS that accumulates sensor data from
IP networks and focuses on the analytics and the mashing up of the
data. The PaaS includes a REST API, data storage functionalities and
OpenId access control for the data. The strength of this platform is
the publish/subscribe mechanism, and querying of data streams, both
from local  and external data sources) to perform analytical tasks.

\iotplatform{IFTTT}{https://ifttt.com/}\label{iot:ifttt} 
(``if this then that'') is a SaaS offering, allowing a rapid
composition of services called ``recipes'' by applying simple if-then
rules to external service building blocks, such as emails, Facebook
events, or Belkin's WeMo switch, that either play the role of a
trigger (if) or an action (then, do). 
Though the service is free to use, the APIs to the service are not
open at the time of writing. The recipes can be personal or shared at
the discrepancy of the user; otherwise, the service building blocks
rather than IFTTT deal with the user generated data. 

\iotplatform{Kahvihub}{http://github.com/uh-cs-iotlab/kahvihub}
\label{iot:kahvihub}
The Kahvihub platform is open-source and designed to be extremely
extendable, as all components in the Kahvihub are delivered by
third-parties, in the form of plugins or applications. 
These components are preferably scripted, to ensure a high re-usability
of the platform's operations on a different platform implementation
(the platform is expected to be deployed on various and heterogeneous
hardware). 
The Kahvihub prototype is aiming to enable edge analytics by creating
local networks of IoT devices that can collaboratively and autonomously
analyze the data that they produce.

\iotplatform{LinkSmart\textsuperscript{\texttrademark}}
{http://www.hydramiddleware.eu/news.php}\label{iot:linksmart}
The LinkSmart\textsuperscript{\texttrademark} middleware platform,
formerly Hydra, is an open-source platform licensed under the
LGPLv3. The platform enable the creation of a network for embedded systems,
using semantics to discover the devices connected to the network. The
middleware is based on a service-oriented architecture. The platform
provides a SDK for application development and a DDK for device
development.

\iotplatform{MyRobots}{http://www.myrobots.com/}
\label{iot:myrobots}
MyRobots is a proprietary cloud-based platform to connect robots
to the IoT. Data format supported are JSON, XML, CSV and the web
services are buildable using REST API. By default, the privacy of
robots is set to public, but can be changed to private. The platform
enables robots to be controlled over the Internet. The platform also
includes an application store.

\iotplatform{Niagara$^{AX}$}{http://www.niagaraax.com/}
\label{iot:niagara}
Niagara\textsuperscript{AX}~\cite{Samad2007} is a  proprietary
M2M dedicated software development framework that is fully
distributed. 
It interconnect heterogeneous devices. However, details are missing
about the nature of the open API.

\iotplatform{Nimbits}{http://www.nimbits.com/}\label{iot:nimbits}
Similarly to \emph{ARM mbed}~\citeiot{iot:armmbed}, the Nimbits
server has been made cloud architecture compatible, hence it scales
from a single private server to a cloud architecture. 
Nimbits includes three levels of privacy for the data: (i) private,
(ii) protected (read-only is public) and (iii) public. Control over
the data and its ownership is to the user. 
The data is transmitted via XMPP messaging protocol. Web services
access the data with HTML POST request and JSON data format. 
The platform is open source licensed under the Apache License 2.0.

\iotplatform{NinjaBlock}{http://ninjablocks.com/}\label{iot:ninjablock}
NinjaBlock provides open-source hardware and open-source software to
facilitate the development of sensors. However, the Ninja platform is
proprietary and cloud-based.
A RESTful API is avaiblable to connect NinjaBlock hardware to the
cloud. NinjaBlock is open-hardware and serves as a gateway between the
sensors and the Ninja platform. 
JSON data format is used by the platform and access is granted via the
OAuth2 authentication protocol.

\iotplatform{Node-RED}{http://nodered.org/}\label{iot:nodered}
Node-RED is an open-source \emph{Node.js} tool that aims to simplify the
connection between IoT devices and web services. It incorporates the
concept of flow for IoT devices and data that allows complex
interactions between objects and services. The flow can be published
on the Node-RED website for sharing. Node-RED is a creation of IBM
Emerging Technology. Some cloud-based services, such as FRED\footnote{
\url{https://fred.sensetecnic.com/}}, provide front-end for Node-RED
and others~\cite{Blackstock2014} integrate Node-RED to their own
platform (e.g., WotKit) for added values.

\iotplatform{OpenIoT}{http://openiot.eu/}\label{iot:openiot}
OpenIoT platform is an open-source platform, fully decentralized, that
provides connectivity with constrained devices such as sensors. The
platform provides a billing mechanism for the use of services.

\iotplatform{OpenMTC}{http://www.open-mtc.org/}\label{iot:openmtc}
Cloud-based solution for M2M that aims to integrate all the standards
defined by the ETSI M2M, oneM2M and 3GPP.

\iotplatform{OpenRemote}{http://www.openremote.org}
\label{iot:openremote}
OpenRemote is a centralized open-source platform, licensed under the
Affero GNU Public License. The platform supports home and domotic
automation spaces using a top-down approach.

\iotplatform{Open.Sen.se}{http://open.sen.se/}\label{iot:opensense}
Open.Sen.se is closed-source PaaS/SaaS. 
A tool called \emph{Funnel} can be used to aggregate data, but only on
data feeds that are within our dashboard. 
It is possible to get the data from different source and mash it
up. The platform uses the JSON data format and REST API for web
services development. 
The privacy of data visualization is either public or private, data is
always private (needs private keys at all times to use the API).

\iotplatform{realTime.io}{https://www.realtime.io/}
\label{iot:realtimeio}
IoBridge realTime.io provides a proprietary cloud-based platform
(PaaS) to connect devices to the Internet and build applications upon
the data. As realTime.io uses a proprietary transport protocol for
data, \emph{ioDP}, the physical devices need to be connected to the
realTime.io cloud service via a proprietary gateway. Once these
gateways are connected to the service, public API (requiring
realTime.io keys) enables the connection to the device to pull or push
data to the devices. 

\iotplatform{SensorCloud\textsuperscript{\texttrademark}}
{http://www.sensorcloud.com/}\label{iot:sensorcloud}
SensorCloud\textsuperscript{\texttrademark} is a proprietary
cloud-based sensor data storage and visualization platform (PaaS).  
It provides a fully REST compliant API and the CSV and XDR data
formats are supported. It also provides tools for visualization and
data mashup (MathEngine). 

\iotplatform{SkySpark}{http://skyfoundry.com/skyspark/}
\label{iot:skyspark}
SkySpark is a proprietary software that can be locally installed on a
private server or on a cloud and enable analytic tools for big data
processing. 
The software does not require the connection of devices to the cloud. 
The software includes a REST API for connection with third-party
applications and web services. 
The SkySpark software does not include direct management of connected
devices.

\iotplatform{Swarm}{http://buglabs.net/products/swarm}
\label{iot:swarm}
Bug's Swarm cloud-based platform (PaaS) is not open-source but
provides an open-source client and some tools (unknown license). 
It creates swarm of resources to consume data, produce data or both
among actors connected to the swarm.
There is limited information on how the swarm data is stored, and who
had its ownership.
A RESTful API and JSON data format are usable to communicate with the
devices. The platforms also provide GUI tools, such an interactive
dashboard with data visualization capabilities.

\iotplatform{TempoDB}{https://tempo-db.com/}\label{iot:tempodb}
TempoDB is a proprietary, cloud-based PaaS that enables the users to
upload their data on the cloud via a REST API.
The service enables to store, retrieve, and query the data, while
ensuring data security, multiple back-ups and providing visualization
tools, etc. 
This service offers billing offers depending on the user need.

\iotplatform{TerraSwarm}{http://www.terraswarm.org/}
\label{iot:terraswarm}
The TerraSwarm project~\cite{Lee2012} envision the development of a
new kind of operating system, the SwarmOS, to natively support the
heterogeneous nature of the devices and solutions existing in the IoT
and enable the infrastructure with the ability to aggregate information
from a variety of data sources. The architecture relies heavily on the
power of cloud computing. The operating system will be also
open-source to improve its reliability and efficiency, while maximizing
the potential of innovative development of ``swarm-apps'' build upon
the system.

\iotplatform{The thing system}{http://thethingsystem.com/}
\label{iot:thethingsystem}
The thing system is a software using \emph{Node.js} that enables
discovery of smart things in the home environment. 
The project is open-source and licensed under the M.I.T license. The
software does not provide storage functionalities and must be coupled
with a PaaS to enable storage outside the home area. 
The software intends only to provide access remotely to smart devices
of smart homes.

\iotplatform{Thing Broker}
{http://www.magic.ubc.ca/wiki/pmwiki.php/ThingBroker/ThingBroker}
\label{iot:thingbroker}
The Thing Broker~\cite{PerezdeAlmeida2013} is a centralized platform
that provides a Twitter-based abstraction model
for \emph{Things} and \emph{Events}, that could be used to create
local ecosystems such as smart homes. A REST API is provided by the
platform to access the data and devices.

\iotplatform{ThingSpeak}{https://www.thingspeak.com/}
\label{iot:thingspeak}
ThingSpeak is decentralized, open-source and copyrighted by ioBridge
under the licence GPLv3. Commercial software or hardware using
ThingSpeak requires a commercial agreement with IoBridge
Inc. ThingSpeak provides a server that may be used to store and
retrieve IoT data. It allows opening of the channels (data flows,
support the JSON, XML, CSV data formats) to the public but do not
provide extensive configuration of the data flows. The platform also
provides visualization tools and enables the creation of widgets in
Javascript/HTML/CSS to visualize the data in a more personified
fashion.

\iotplatform{ThingSquare}{http://thingsquare.com/}
\label{iot:thingsquare}
ThingSquare is a proprietary cloud-based platform specialized on
connecting constrained devices. It require a gateway, but its firmware
is open source. The gateway creates a wireless mesh networks of
sensors and connect it to the Internet. The devices can access the
Internet, but the devices are invisible from outside the mesh. The
platform also includes a protocol for constrained devices. 

\iotplatform{ThingWorx}{http://www.thingworx.com/}
\label{iot:thingworx}
ThingWorx is a proprietary cloud-based M2M dedicated platform (PaaS).
It provides a variety of tools and services to support end-to-end
solutions. The devices and data are accessible via a REST API. 
Due to acquisition of Axeda\footnote{\url{http://www.thingworx.com/news/\\ptc-to-acquire-axeda-to-expand-internet-of-things-technology-portfolio}}, 
the platform will likely be expanded with the IoT connectivity services, 
software agents and toolkits of the latter, 
Axeda being a proprietary cloud-based solution for M2M communication of
businesses and one of the key player in the current IoT landscape.

\iotplatform{Sense Tecnic WoTkit}{http://sensetecnic.com/}
\label{iot:wotkit}
The WoTkit~\cite{Lea2014a} is a proprietary cloud-based platform
that offers an interesting search tool for public sensor. Public
sensors do not require an account to be used.

\iotplatform{Xively}{https://xively.com/}\label{iot:xively}
Xively (formerly Pachube) is a proprietary cloud-based platform (PaaS).
Ownership of the data remains to the user, but the data is stored on
the Xively server. 
Xively provides open-source APIs (in various programming languages)
mostly with the BSD 3-clause license.
Xively provides an extensive RESTful API including a search tool in
order to retrieve feeds (flow of data) depending on selected
characteristics (location radius, name, type of data stored, etc.)

\bibliographystyle{elsarticle-num}
\bibliography{iot-platforms-gap-analysis}

\end{document}